\documentclass[aps,prl,onecolumn,showpacs,preprintnumbers]{revtex4}
\usepackage{amsmath}
\usepackage{dcolumn}
\usepackage{bm}
\usepackage{subfigure}
\usepackage{amsfonts}
\usepackage{amssymb}
\usepackage{makeidx}
\usepackage{epsfig}
\usepackage{graphicx}

\begin{document}

\title{Hamiltonian approach to GR - Part 1: covariant theory of classical
gravity}
\author{Claudio Cremaschini}
\affiliation{Institute of Physics and Research Center for Theoretical Physics and
Astrophysics, Faculty of Philosophy and Science, Silesian University in
Opava, Bezru\v{c}ovo n\'{a}m.13, CZ-74601 Opava, Czech Republic}
\author{Massimo Tessarotto}
\affiliation{Department of Mathematics and Geosciences, University of Trieste, Via
Valerio 12, 34127 Trieste, Italy}
\affiliation{Institute of Physics, Faculty of Philosophy and Science, Silesian University
in Opava, Bezru\v{c}ovo n\'{a}m.13, CZ-74601 Opava, Czech Republic}
\date{\today }

\begin{abstract}
A challenging issue in General Relativity concerns the determination of the
manifestly-covariant continuum Hamiltonian structure underlying the Einstein
field equations and the related formulation of the corresponding covariant
Hamilton-Jacobi theory. The task is achieved by adopting a synchronous
variational principle requiring distinction between the prescribed
deterministic metric tensor $\widehat{g}(r)\equiv \left\{ \widehat{g}_{\mu
\nu }(r)\right\} $\ solution of the Einstein field equations which
determines the geometry of the background space-time and suitable
variational fields $x\equiv \left\{ g,\pi \right\} $ obeying an appropriate
set of continuum Hamilton equations, referred to here as GR-Hamilton
equations$.$ It is shown that a prerequisite for reaching such a goal is
that of casting the same equations in evolutionary form by means of a
Lagrangian parametrization for a suitably-reduced canonical state. As a
result, the corresponding Hamilton-Jacobi theory is established in
manifestly-covariant form. Physical implications of the theory are
discussed. These include the investigation of the structural stability of
the GR-Hamilton equations with respect to vacuum solutions of the Einstein
equations, assuming that wave-like perturbations are governed by the
canonical evolution equations.
\end{abstract}

\maketitle

\section{1 - Introduction}

This is the first paper of a two-part investigation dealing with the
Hamiltonian theory of the gravitational field, and more precisely, the one
which is associated with the so-called Standard Formulation of General
Relativity (SF-GR) \cite{ein1,LL,gravi,wald}, i.e., the Einstein field
equations. In the second paper the corresponding quantum formulation will be
presented. For this purpose, in the two papers new manifestly-covariant
variational approaches are developed. The two formulations will be referred
to as theories of\ Covariant Classical and respectively Quantum Gravity
(CCG/CQG) or briefly \emph{CCG- }and\emph{\ CQG-theory}.

As shown below \emph{CCG-theory} is built upon the results presented in Refs.%
\cite{noi1,noi2} about the variational formulation of GR achieved in the
context of a\textbf{\ }\emph{DeDonder-Weyl}\textbf{-}type approach and the
corresponding possible realization of a super-dimensional and\textbf{\ }%
manifestly-covariant Hamiltonian theory. In particular, based on a suitable
identification of the effective kinetic energy and the related Hamiltonian
density $4-$scalars adopted in Ref.\cite{noi2}, the aim here is to identify
a reduced-dimensional continuum Hamiltonian structure of SF-GR,\ to be
referred to here as \emph{Classical Hamiltonian Structure} (CHS). The
crucial goal of the paper is to show that CHS can be associated with
arbitrary possible solutions of the Einstein field equations corresponding
either to vacuum or non-vacuum conditions. In other words, this means that
the same Hamiltonian structure should occur for arbitrary external source
terms which may appear in the variational potential density.

Despite being intimately related to the one earlier considered in Ref.\cite%
{noi2}, the new Hamiltonian structure is achieved in fact by means of the
parametrization of the corresponding canonical state in terms of the
proper-time determined along arbitrary geodetics of the background metric
field tensor. This feature turns out to be of paramount importance for the
establishment of the corresponding canonical transformation and covariant
Hamilton-Jacobi theories. The CHS\textbf{\ }determined in this way is shown
to be realized by the ensemble $\left\{ x_{R},H_{R}\right\} $ represented by
an appropriate variational canonical state $x_{R}=\left\{ g,\pi \right\} $,\
with $g$\ and $\pi $\ being suitably-identified tensor fields representing
appropriate continuum Lagrangian coordinates and conjugate momenta and $%
H_{R} $ a corresponding variational Hamiltonian density.

Its basic feature is that of being based, in analogy with Ref.\cite{noi2},
on the adoption of the \emph{synchronous Hamiltonian variational principle}
for the variational formulation of GR. However, in difference to the same
reference, new features are added which, as we intend to show, are mandatory
for the construction of the corresponding canonical transformation and
Hamilton-Jacobi theories. In particular for this purpose, first, a
reduced-dimensional representation is introduced for the canonical momenta,
which however leaves formally unchanged the corresponding variational
Hamiltonian density. Second, an appropriate parametrization by means of a
suitably-defined proper time $s$\ is introduced so that the resulting
Euler-Lagrange equations are now realized by means of Hamilton equations in
evolution form.

Accordingly, variational and prescribed tensor fields\ are introduced, with
the prescribed ones being left invariant by the synchronous variations and
the variational fields. In particular, the same Hamilton equations may
reduce identically to the Einstein field equations, which are fulfilled by
the prescribed fields, if suitable initial conditions are set.\textbf{\ }%
This occurs provided the Poisson bracket of the Hamiltonian density is a
local function, i.e., it does not depend explicitly on proper time.\ In the
realm of the classical theory the physical behavior of variational fields
provide the mathematical background for the establishment of a
manifestly-covariant Hamiltonian theory of GR, and in particular the
CCG-theory realized here. When passing to the corresponding covariant
quantum theory (i.e., in the present case the CQG-theory to be developed in
the subsequent paper) variational fields become\ quantum fields and inherit
the corresponding tensor transformation laws of classical fields. Thanks to
its intrinsic consistency with the principles of covariance and manifest
covariance,\ the synchronous variational setting developed in Refs.\cite%
{noi1,noi2} provides at the same time:\newline
- the natural framework for a Hamiltonian theory of classical gravity which
is consistent with SF-GR;\newline
- the prerequisite for the establishment of a covariant quantum theory of
gravitational field which is in turn consistent with classical theory and
SF-GR (see Ref.\cite{part2}, hereon Part 2).

According to such an approach the $4-$scalar, \textit{i.e., }invariant, $4-$%
volume element of the space-time ($d\Omega $) entering the action
functional\ is considered independent of the functional class of variations,
so that it must be defined in terms of a \emph{prescribed metric tensor field%
} $\widehat{g}\left( r\right) $, represented equivalently either in terms of
its covariant or counter variant component, \textit{i.e.,} either\textbf{\ }$%
\widehat{g}\left( r\right) \equiv \left\{ \widehat{g}_{\mu \nu }\left(
r\right) \right\} $\ or\textbf{\ }$\widehat{g}\left( r\right) \equiv \left\{
\widehat{g}^{\mu \nu }\left( r\right) \right\} $.\textbf{\ }Here $r\equiv
\left\{ r^{\mu }\right\} $ and $\widehat{g}\left( r\right) $ denote
respectively an arbitrary GR-frame parametrization and an arbitrary
particular solution of the Einstein field equations.\textbf{\ }This is
obtained therefore upon identifying in the action functional $d\Omega \equiv
d^{4}r\sqrt{-\left\vert \widehat{g}\right\vert }$, with $d^{4}r$ being%
\textbf{\ }the corresponding canonical measure expressed in terms of the
said parametrization and $\left\vert \widehat{g}\right\vert $ denoting as
usual the determinant of the metric tensor $\widehat{g}\left( r\right) $. In
the context of the synchronous variational principle to GR a further
requirement is actually included which demands that the prescribed field\ $%
\widehat{g}_{\mu \nu }(r)$ must determine, besides $d\Omega $,\ also the
\emph{geometric properties} of space-time. This means that $\widehat{g}_{\mu
\nu }(r)$ should uniquely prescribe the tensor transformation laws of
arbitrary tensor fields, which may depend in principle, besides $\widehat{g}%
_{\mu \nu }(r)$,\ both on the variational state $x_{R}$ and the $4-$position
$r\equiv \left\{ r^{\mu }\right\} $. This requires in particular that $%
\widehat{g}_{\mu \nu }(r)$\ and $\widehat{g}^{\mu \nu }(r)$\ respectively
lower and raise tensor indexes of the same tensor fields. In a similar way $%
\widehat{g}_{\mu \nu }(r)$ uniquely determines also the standard Christoffel
connections which enter both the Ricci tensor $\widehat{R}_{\mu \nu }$ and
the covariant derivatives of arbitrary variational tensor fields. Therefore,
in the context of synchronous variational principle to GR the approach known
in the literature as "background space-time picture"\ \cite%
{Drummond2000,Moffayt2012,Hossenfelder2009}\textbf{\ }is adopted, whereby
the background space-time\textbf{\ }$\left( \mathbf{Q}^{4},\widehat{g}\left(
r\right) \right) $\textbf{\ }is considered defined \textquotedblleft a
priori\textquotedblright\ in terms of $\widehat{g}_{\mu \nu }(r)$,\ while
leaving unconstrained all the variational fields $x_{R}=\left\{ g,\pi
\right\} $\ and in particular the Lagrangian coordinates $g\left( r\right)
\equiv \left\{ g_{\mu \nu }\left( r\right) \right\} $.

Indeed, consistent with Ref.\cite{noi2}, the physical interpretation which
arises from CCG-theory exhibits a connection also with the so-called \textit{%
induced gravity} (or \textit{emergent gravity}) \cite{emerg2,emerg1}, namely
the conjecture that the geometrical properties of space-time should reveal
themselves as a mean field description of microscopic stochastic or quantum
degrees of freedom underlying the classical solution. In the present
approach this is achieved by introducing the prescribed metric tensor $%
\widehat{g}_{\mu \nu }\left( r\right) $ in the Lagrangian and Hamiltonian
action functionals, which is held constant in the variational principles
when performing synchronous variations and has to be distinguished from the
variational field $g_{\mu \nu }\left( r\right) $. In this picture, $\widehat{%
g}_{\mu \nu }\left( r\right) $ should arise as a macroscopic prescribed mean
field emerging from a background of variational fields $g_{\mu \nu }\left(
r\right) $, all belonging to a suitable functional class. This permits to
introduce a new representation for the action functional in superabundant
variables, depending both on $g_{\mu \nu }\left( r\right) $ and $\widehat{g}%
_{\mu \nu }\left( r\right) $. Such a feature, as explained above, is found
to be instrumental for the identification of the covariant Hamiltonian
structure associated with the classical gravitational field and provides a
promising physical scenario where to develop a covariant quantum treatment
of GR.

In this reference, one has to acknowledge the fact that the Hamiltonian
description of classical systems is a mandatory conceptual prerequisite for
achieving a corresponding quantum description \cite{FoP1,FoP2}, \textit{%
i.e., }in the case of continuum systems, the related relativistic quantum
field theory. This task involves the identification of the appropriate
Hamiltonian representation of the continuum field, to be realized by means
of the following steps:\newline
\emph{Step \#1: }Establishment of underlying Lagrangian and Hamiltonian
variational action principles.\newline
\emph{Step \#2: }Construction of the corresponding Euler-Lagrange equations,%
\emph{\ }realized respectively in terms of appropriate continuum Lagrangian
and Hamiltonian equations.\newline
\emph{Step \#3: }Determination of the corresponding set of continuum
canonical transformations and\ formulation of the related Hamilton-Jacobi
theory.

The proper realization of these steps remains crucial. In actual fact, the
last target appears as a prerequisite of foremost importance for being able
to reach a consistent formulation of relativistic quantum field theory for
General Relativity, \textit{i.e., }the so-called Quantum Gravity. The
conclusion follows by analogy with Electrodynamics. In fact, as it emerges
from the recent investigation concerning the axiomatic foundations of
relativistic quantum field theory for the radiation-reaction problem
associated with classical relativistic extended charged particles (see Refs.%
\cite{EPJ1,EPJ2,EPJ3,EPJ5,EPJ7,EPJ8}), it is the Hamilton-Jacobi theory
which naturally provides the formal axiomatic connection between classical
and quantum theory, to be established by means of a suitable realization of
the quantum correspondence principle.

Prerequisite for reaching such goals in the context of relativistic quantum
field theory is the establishment of a theory fulfilling at all levels both
the \emph{Einstein general covariance principle} and the \emph{principle of
manifest covariance}. Such a viewpoint is mandatory in order that the
axiomatic construction method of SF-GR makes any sense at all \cite{haw}.
Indeed, in order that physical laws have an objective physical character
they cannot depend on the choice of the GR reference frame. This requisite
can only be met provided all classical physical observables and the
corresponding mathematical relationships holding among them, \textit{i.e., }%
the physical laws, can actually be expressed in tensorial form with respect
to the group of transformations indicated above. In the context of SF-GR the
adoption of the same strategy requires therefore the realization of \emph{%
Steps \#1-\#3\ }in manifest covariant form. As far as the actual
identification of \emph{Steps \#1 }and\emph{\ \#2 }for SF-GR is concerned,
the candidate is represented by the variational theory reported in Refs.\cite%
{noi1,noi2}. The distinctive features of such a variational theory, which
sets it apart from previous Hamiltonian formulations in literature \cite%
{h1,h3,h4,h5,h6,h7}, lie in its consistency with the criteria indicated
above and the DeDonder-Weyl classical field theory approach \cite%
{donder,weyl,sym3,sym4,sym5,sym6,sym7,sym8,sym9}.

Nevertheless well-known alternative approaches exist in the literature which
are based on non-manifestly covariant approaches.\ For the purpose of formal
comparison let us briefly mention some of them, a detail analysis being left
to future developments. For definiteness, approaches can be considered which
are built upon space-time\ foliations, namely\ are based on so-called 3+1
and/or 2+2 splitting schemes.\ In fact, GR can be formulated in any GR-frame
(i.e., coordinate system) by introducing a suitable local point
transformation $r^{\mu }\rightarrow r^{\prime \mu }=f^{\mu }(r)$ leading to
a decomposition of this type. In particular, the 3+1 approach is convenient
for purposes related, for example, to the definition of conventional
energy-momentum tensors, thermodynamic and kinetic values,\ and to provide
corresponding methods of quantization \cite{zzz1,zzz2,zzz3}. The latter are\
exemplified by the well-known approach developed by Arnowitt, Deser and
Misner (1959-1962 \cite{ADM}), usually referred to as ADM theory in
literature. The same\ theory is based on the introduction of the so-called
3+1 decomposition of space-time which by construction is foliation
dependent, in the sense that it relies on a peculiar choice of a family of
GR frames for which time and space transform separately so that space-time
is effectively split into the direct product of a 1-dimensional time and a
3-dimensional space subsets respectively (ADM-foliation) \cite{alcu}.
Instead, different types of 2+2 splitting (or with double 3+1 and 2+2
splitting) are considered, for instance, to find new classes of GR exact
solutions \cite{Vaca2,Vaca1,Vaca3}, to develop the theory of geometric flows
related to classical, quantum gravity and geometric thermodynamics \cite%
{Cli,Vaca6}, or to elaborate some approaches based on deformation
quantization of GR and modified gravity theories \cite{Vaca4,Vaca5}. In
comparison with these approaches, the manifestly-covariant Lagrangian and
Hamiltonian formulations of GR reported in Refs.\cite{noi1,noi2} and
developed below mainly differ because, first, there is no introduction of
foliation of space-time, so that the 4-tensor formalism is preserved at all
stages of investigation. Second, in contrast to the Hamiltonian theory of GR
obtained from ADM decomposition \cite{wald}, both Lagrangian and Hamiltonian
dynamical variables and canonical state are expressed in 4-tensor notation
and satisfy as well the manifest covariance principle.\ Third, in the
context of CCG-theory\ the Hamiltonian flow\ associated with the Hamiltonian
structure\ $\left\{ x_{R},H_{R}\right\} $\ (see Eq.(\ref{flow}) below) is
defined with respect to an invariant proper-time $s$, and not a\
coordinate-time as in ADM theory. Finally, it must be stressed that, in such
a context\ for the proper implementation of the DeDonder-Weyl formalism,
besides the customary 4-scalar curvature term of the Einstein-Hilbert
Lagrangian, $4-$tensor (i.e., manifestly-covariant)\ momenta must be\
adopted in the action functional. This property which can be fulfilled only\
adopting a synchronous variational principle is\ missing in the ADM
Hamiltonian theory, where field variables and conjugate momenta are
identified only after performing the 3+1 foliation on the Einstein-Hilbert
Lagrangian of the associated asynchronous variational principle. Despite
this difference the two approaches are\ complementary in the sense that they
exhibit distinctive physical properties associated with the\ two canonical
Hamiltonian structures underlying SF-GR\ (see again Ref.\cite{noi2}).

\subsection{Goals of the paper}

Going beyond the considerations discussed above, the construction of
CCG-theory involves a number of questions, closely related to the continuum
Hamiltonian theory reported in Ref.\cite{noi2}, which remain to be
addressed. This involves posing the following distinct goals:

\begin{itemize}
\item \emph{GOAL \#1: Reduced continuum Hamiltonian theory for SF--GR -} The
search for a reduced-dimensional realization of the continuum Hamiltonian
theory for the Einstein field equations, which still satisfies the principle
of manifest covariance. In fact, as a characteristic feature of the
DeDonder-Weyl approach, in the Hamiltonian theory given in Ref.\cite{noi2}
the canonical variables defining the canonical state have different
tensorial orders, with the momenta being realized by third-order $4-$%
tensors. In contrast, the new approach to be envisaged here should provide a
realization of the canonical state $x_{R}\equiv \left\{ g_{\mu \nu },\pi
_{\mu \nu }\right\} $ in which both generalized coordinates and
corresponding momenta have the same tensorial dimension and are represented
by second-order $4-$tensor fields.

\item \emph{GOAL \#2: Evolution form of the reduced continuum Hamilton
equations - }A further problem is whether the same reduced continuum
Hamilton equations can be given a causal evolution form, namely they can be
cast as \emph{canonical evolution equations}. Since originally the continuum
Hamilton equations are realized by PDE, this means that some sort of
Lagrangian representation should be determined. Hence, by introducing a
suitable Lagrangian Path (LP) parametrization of\ the canonical state $%
x_{R}\equiv \left\{ g_{\mu \nu },\pi _{\mu \nu }\right\} $ in terms of the
proper-time associated with the prescribed tensor field $\widehat{g}_{\mu
\nu }(r)$ indicated above, the corresponding continuum canonical equations
are found to be realized by means of evolution equations advancing in proper
time the canonical state. These will be referred to as \emph{GR-Hamilton
equations}\textbf{\ }of CCG-theory: they generate the evolution of the
corresponding canonical fields by means of a suitable canonical flow.

\item \emph{GOAL \#3: Realization of manifestly-covariant continuum
Hamilton-Jacobi theory -} A related question which arises involves, in
particular, the determination of the canonical transformation which
generates the flow corresponding to the continuum canonical evolution
equations. This concerns, more precisely, the development of a corresponding
Hamilton-Jacobi theory\ applicable in the context of CCG-theory and the
investigation of the canonical transformation generated by the corresponding
Hamilton principal function.

\item \emph{GOAL \#4:\ Global prescription and regularity properties of the
corresponding GR-Lagrangian and Hamiltonian densities}. The Lagrangian and
Hamiltonian formulations should be globally prescribed in the appropriate
phase-spaces. The global prescription should include also the validity of
suitable \emph{regularity properties} of the corresponding Hamiltonian
density $H_{R}$.

\item \emph{GOAL \#5:\ Identification of the gauge properties of the
classical GR-Lagrangian and Hamiltonian densities. }The related issue
concerns the identification of the possible gauge indeterminacies, in terms
of suitable \emph{gauge functions}, characterizing the Lagrangian and
Hamiltonian densities.

\item \emph{GOAL \#6: Dimensionally-normalized form of CHS.} In particular,
the\ goal here is to show that a suitable\ dimensional normalization of the
Hamiltonian structure $\left\{ x_{R},H_{R}\right\} $ can be reached so that
the canonical momenta acquire the physical dimensions of an action, a
feature required for the establishment of a quantum theory of GR in terms of
Hamilton-Jacobi theory. More precisely this involves\ the construction of a
non-symplectic canonical transformation for the GR-Hamiltonian density $%
H_{R} $. The issue is to show that this can be taken of the form:%
\begin{equation}
\left\{
\begin{array}{c}
g^{\mu \nu }\rightarrow \overline{g}^{\mu \nu }=g^{\mu \nu } \\
\pi ^{\mu \nu }\rightarrow \overline{\pi }^{\mu \nu }=\frac{\alpha L}{k}\pi
^{\mu \nu } \\
H_{R}\rightarrow \overline{H}_{R}\equiv \overline{T}_{R}+\overline{V}=\frac{%
\alpha L}{k}H_{R}%
\end{array}%
\right. ,  \label{CANONICAL-0}
\end{equation}%
where $\kappa $ is the dimensional constant $\kappa =\frac{c^{3}}{16\pi G}$,
$L$ is a $4-$scalar scale length to be defined, $\alpha $ is a suitable
dimensional $4-$scalar, while $\overline{T}_{R},$ and $\overline{V}$ denote
the corresponding transformed effective kinetic and potential densities
defining the transformed Hamiltonian density $\overline{H}_{R}$. Then the
question arises whether $\alpha $ can be prescribed in such a way that the
transformed canonical momentum $\overline{\pi }^{\mu \nu }$ has the
dimensions of an action.

\item \emph{GOAL \#7: Structural stability of the GR-Hamilton equations of
CCG-theory -} The final issue concerns the study of the structural stability
which in the framework of CCG-theory the canonical equations\ exhibit with
respect to their stationary solutions, \textit{i.e., }the solutions of the
Einstein equations. In fact, depending on the specific realization of CHS
considered here, infinitesimal perturbations whose dynamics is governed by
the said canonical evolution equations may exhibit different stability
behaviors, \textit{i.e., }be stable/unstable or marginally stable, with
respect to arbitrary solutions of the Einstein field equations. For
definiteness, the case of vacuum solutions with a non-vanishing cosmological
constant $\Lambda $ is treated. It is shown that the stability analysis
provides a prescriptions for the gauge functions indicated above which
characterize the GR-Hamiltonian density.
\end{itemize}

In view of these considerations and of the results already achieved in\ Refs.%
\cite{noi1,noi2}, in this paper the attention will be focused on the
investigation of \emph{GOALS \#1-\#7}. These topics, together with the
continuum Lagrangian and Hamiltonian theories proposed in Refs.\cite%
{noi1,noi2}, have potential impact in the context of both classical and
quantum theories of General Relativity.

\section{2 - Evolution form of Hamilton equations for SF-GR}

In this section the problem of\ the determination of a\textbf{\ }\emph{%
reduced continuum Hamiltonian theory }for GR is addressed for a prescribed%
\textbf{\ }Hamiltonian system. This is represented by the CHS $\left\{
x_{R},H_{R}\right\} $\ which is formed by an appropriate $4-$tensor
canonical state $x_{R}$\ and a suitable $4-$scalar Hamiltonian density $%
H_{R}\left( x_{R},\widehat{x}_{R}(r),r,s\right) $. In particular, the target
requires to find a realization of the variational canonical momentum in such
a way that, in the corresponding reduced canonical state, fields and reduced
momenta form a couple of second-rank conjugate $4-$tensors. The requisite is
that such a Hamiltonian theory should warrant the validity of the non-vacuum
Einstein field equations to be achieved as realizations of suitable reduced
continuum Hamilton equations set in evolution form and referred to as\emph{\
GR-Hamilton equations }of CCG-theory.\ More precisely, these are realized by
the initial-value problem represented by the canonical equations:
\begin{equation}
\left\{
\begin{array}{c}
\frac{Dg_{\mu \nu }(s)}{Ds}=\frac{\partial H_{R}(x_{R},\widehat{x}%
_{R}(r),r,s)}{\partial \pi ^{\mu \nu }(s)}, \\
\frac{D\pi _{\mu \nu }(s)}{Ds}=-\frac{\partial H_{R}\left( x_{R},\widehat{x}%
_{R}(r),r,s\right) }{\partial g^{\mu \nu }(s)},%
\end{array}%
\right.  \label{canonical evolution equations -2}
\end{equation}%
and the initial conditions of the type
\begin{equation}
\left\{
\begin{array}{c}
g_{\mu \nu }(s_{o})\equiv g_{\mu \nu }^{(o)}(s_{o}), \\
\pi _{\mu \nu }(s_{o})\equiv \pi _{\mu \nu }^{(o)}(s_{o}).%
\end{array}%
\right.  \label{initial conditions}
\end{equation}%
Then the solution of the initial-value problem (\ref{canonical evolution
equations -2})-(\ref{initial conditions}) generates the Hamiltonian flow%
\begin{equation}
x_{R}(s_{o})\rightarrow x_{R}(s),  \label{flow}
\end{equation}%
which is associated with the Hamiltonian structure\emph{\ }$\left\{
x_{R},H_{R}\right\} .$ Here the notation is as follows. First, $s$\ denotes
the proper time prescribed along an arbitrary geodesic curve $r(s)\equiv
\left\{ r^{\mu }(s)\right\} $. This is associated with the prescribed metric
tensor $\widehat{g}_{\mu \nu }(r)$\ of the background space-time. Second,%
\begin{equation}
x_{R}(s)\equiv \left\{ g_{\mu \nu }(r(s)),\pi _{\mu \nu }(r(s))\right\}
\label{REDUCED STATE}
\end{equation}%
identifies the $s-$parametrized \emph{reduced-dimensional variational
canonical state}, with $g_{\mu \nu }(r)$\ and $\pi _{\mu \nu }(r)$\ being
the corresponding continuum Lagrangian coordinates and the conjugate
momenta, $\widehat{x}_{R}(s)\equiv \left\{ \widehat{g}_{\mu \nu }(r(s)),%
\widehat{\pi }_{\mu \nu }(r(s))\equiv 0\right\} $\textbf{\ }being the
corresponding prescribed state and $H_{R}(x_{R},\widehat{x}_{R}(r),r,s)$\
the variational Hamiltonian $4-$scalar density to be suitably determined.
Finally, $\frac{D}{Ds}$\ is the covariant $s-$derivative%
\begin{equation}
\frac{D}{Ds}=\frac{\partial }{\partial s}+t^{\alpha }(s)\widehat{\nabla }%
_{\alpha },  \label{covariant s-derivative}
\end{equation}%
while $t^{\alpha }(s)$\ and $\widehat{\nabla }_{\alpha }$\ are respectively
the tangent $4-$vector to the geodesics $r(s)\equiv \left\{ r^{\mu
}(s)\right\} $\ and the covariant derivative evaluated at the same position
in terms of the prescribed metric tensor $\widehat{g}_{\mu \nu }(r)$.

The GR-Hamilton equations are covariant with respect to arbitrary canonical
transformations. This property implies that the same equations are covariant
with respect to an arbitrary local point transformation (LPT) which leaves
invariant a given space-time represented by the differential manifold of the
type $\left( \mathbf{Q}^{4},\widehat{g}(r)\right) $, so that the General
Covariance Principle and the Principle of Manifest Covariance are
necessarily both fulfilled by construction.

The realization of the evolution form of the GR-Hamilton equations
represents a requirement for the construction of a corresponding
manifestly-covariant Hamilton-Jacobi theory of GR. The construction of these
equations is based on the following steps.

\subsection{2A - Step \#1: Prescription of the reduced-dimensional
Hamiltonian density}

In the first step the Hamiltonian density $H_{R}(x_{R},\widehat{x}%
_{R}(r),r,s)$ is identified, extending the treatment given in Ref.\cite{noi2}%
. In terms of the reduced canonical variables this yields%
\begin{equation}
H_{R}\left( x,\widehat{x},r\right) \equiv T_{R}\left( x_{R},\widehat{x}%
_{R}\right) +V(g,\widehat{x},r),  \label{reduced HAMILTONIAN}
\end{equation}%
where the effective kinetic and potential densities $T_{R}\left( x_{R},%
\widehat{x}_{R}\right) $ and $V(g,\widehat{x},r,s)$ can be taken
respectively of the general form%
\begin{equation}
\left\{
\begin{array}{c}
T_{R}\left( x_{R},\widehat{x}_{R}\right) \equiv \frac{1}{2\kappa f(h)}\pi
_{\mu \nu }\pi ^{\mu \nu }, \\
V\left( g,\widehat{x},r,s\right) \equiv \sigma V_{o}\left( g,\widehat{x}%
\right) +\sigma V_{F}\left( g,\widehat{x},r,s\right) .%
\end{array}%
\right.  \label{KINETIC ENERGY DENSITY}
\end{equation}%
Here $f(h)$\ and $\sigma $\ denote suitable \emph{multiplicative gauge }%
functions which remain in principle still arbitrary at this point.\textbf{\ }%
More precisely, $f(h)$\ identifies an \textquotedblleft a
priori\textquotedblright\ arbitrary non-vanishing and
smoothly-differentiable real gauge function depending on the variational
weight-factor $h=\left( 2-\frac{1}{4}g^{\alpha \beta }g_{\alpha \beta
}\right) $\ introduced in Ref.\cite{noi1} and prescribed in such a way that
\begin{equation}
f(\widehat{g}^{\mu \nu }(r))=1.  \label{COINSTRAIUNT ON f(h)}
\end{equation}%
We anticipate here that the function $f(h)$ will be shown in Part 2 to be
identically $f(h)=1$, as required by quantum theory of GR. Furthermore, $%
\sigma $ denotes the additional constant gauge function $\sigma =\pm 1.$
Finally, in Eq.(\ref{KINETIC ENERGY DENSITY}) the two $4-$scalars
\begin{equation}
\begin{array}{c}
V_{o}\left( g,\widehat{x}\right) \equiv \kappa h\left[ g^{\mu \nu }\widehat{R%
}_{\mu \nu }-2\Lambda \right] , \\
V_{F}\left( g,\widehat{x},r\right) \equiv hL_{F}\left( g,\widehat{x}%
,r\right) ,%
\end{array}
\label{POT-ENERGY-SOURCES-2}
\end{equation}%
identify respectively the gravitational and external-field source
contributions defined in Ref.\cite{noi1}, with $L_{F}$ being associated with
a non-vanishing stress-energy tensor.

\subsection{2B - Step \#2: Lagrangian path parametrization}

In the second step we introduce the notion of Lagrangian path (LP) \cite%
{FoP1,FoP2}. \ For this purpose, preliminarily the real $4-$tensor $%
t^{\gamma }(\widehat{g}(r),r)$ is introduced such that identically%
\begin{equation}
\left\{
\begin{array}{c}
t^{\alpha }(\widehat{g}(r),r)\widehat{\nabla }_{\alpha }t^{\gamma }(\widehat{%
g}(r),r)=0, \\
\widehat{g}_{\gamma \delta }(r)t^{\gamma }(\widehat{g}(r),r)t^{\delta }(%
\widehat{g}(r),r)=1,%
\end{array}%
\right.  \label{REQUIREMENT-A}
\end{equation}%
so that by construction $t^{\gamma }(\widehat{g}(r),r)$ is tangent to an
arbitrary geodetics belonging to an arbitrary $4-$position $r\equiv \left\{
r^{\mu }\right\} $ of the space-time $\left( \mathbf{Q}^{4},\widehat{g}%
(r)\right) $ \cite{LL}. Then, the LP is identified with the geodetic curve%
\begin{equation}
\left\{ r^{\mu }(s)\right\} \equiv \left\{ \left. r^{\mu }(s)\right\vert
\text{ }\forall s\in
\mathbb{R}
,\text{ }r^{\mu }(s_{o})=r_{o}^{\mu }\right\} ,
\end{equation}%
which is solution of the initial-value problem%
\begin{equation}
\left\{
\begin{array}{c}
\frac{dr^{\mu }(s)}{ds}=t^{\mu }(s), \\
r^{\mu }(s_{o})=r_{o}^{\mu }.%
\end{array}%
\right.  \label{LP equation}
\end{equation}%
Here the $4-$scalar proper-time $s$ is defined along the same curve $\left\{
r^{\mu }(s)\right\} $ so that $ds^{2}=\widehat{g}_{\mu \nu }(r)dr^{\mu
}(s)dr^{\nu }(s)$. Furthermore, $t^{\mu }(s)$ identifies the $LP-$%
parametrized $4-$vector $t^{\mu }(s)\equiv t^{\mu }(\widehat{g}(r(s)),r(s))$%
. In Eq.(\ref{LP equation}) $\frac{d}{ds}\equiv \frac{\partial }{\partial s}$%
\ identifies the ordinary derivative with respect to $s$.\textbf{\ }In the
following we shall denote as \emph{implicit }$s-$\emph{dependences }the
dependences on the proper time $s$\ appearing in the variational fields
through the LP parametrization of the fields. In contrast, we shall denote
as \emph{explicit }$\emph{s-}$\emph{dependences}\textbf{\emph{\ }}the
proper-time dependences which enter either explicitly on $s$ itself or
through the dependence on $r(s)\equiv \left\{ r^{\mu }(s)\right\} $.

Let us now introduce the parametrization obtained replacing everywhere, in
all the relevant tensor fields, $r\equiv \left\{ r^{\mu }\right\} $ with $%
r(s)\equiv \left\{ r^{\mu }(s)\right\} $, namely obtained identifying
\begin{equation}
\left\{
\begin{array}{c}
g^{\mu \nu }(s)\equiv g^{\mu \nu }(r(s)), \\
\pi ^{\mu \nu }(s)\equiv \pi ^{\mu \nu }(r(s)), \\
\widehat{x}(r)\equiv \widehat{x}(r(s)).%
\end{array}%
\right.
\end{equation}%
This yields for the Hamiltonian density $H_{R}$\ the so-called
LP-parametrization, in terms of which the reduced Hamilton equations (\ref%
{canonical evolution equations -2}) can in turn be represented.\textbf{\ }In
the remainder, for greater generality, such a representation shall be taken
of the form%
\begin{equation}
H_{R}(s)\equiv H_{R}\left( x_{R}(s),\widehat{x}_{R}(r),r(s),s\right) ,
\label{LP-PARAMETRIZATION OF H_R}
\end{equation}%
\textit{i.e., }including also a possible explicit dependence in terms of the
proper time $s$. Specific examples in which explicit $s-$dependences may
occur in the theory include:

1) Continuum canonical transformations and in particular canonical
transformations generating local or nor local point transformations (see Ref.%
\cite{noi4}). In this case explicit $s-$dependences may arise in the
transformed Hamiltonian density due to explicit $s-$dependent generating
functions.

2) Hamilton-Jacobi theory (see Section 3), where in a similar way the\
explicit $s-$dependence in the Hamiltonian density may be generated by the
canonical flow.

3) Stability theory for wave-like perturbations where explicit $s-$%
dependences may appear in the variational fields $x_{R}=x_{R}(s)$\ (see
Section 5).

\subsection{2C - Step \#3: The reduced Hamiltonian variational principle}

Given these premises, in the context of CCG-theory the explicit construction
of the GR-Hamilton equations (\ref{canonical evolution equations -2})
follows in analogy with the extended Hamiltonian theory achieved in Ref.\cite%
{noi2}.\ The goal also in the present case is in fact the development of a
manifestly-covariant variational approach, \textit{i.e.}, in which at all
levels all variational fields, including the canonical variables, the
Hamiltonian density, as well as their synchronous variations and the related
Euler-Lagrange equations, are expressed in $4-$tensor form. To this end in
the framework of the synchronous variational principle developed there - and
in agreement with the DeDonder-Weyl approach - the variational functional is
identified with a real $4-$scalar%
\begin{equation}
S_{R}\left( x,\widehat{x}\right) \equiv \int d\Omega L_{R}\left( x,\widehat{x%
},r,s\right) ,  \label{FUNCTIONAL}
\end{equation}%
with $L_{R}\left( x,\widehat{x},r,s\right) $ being the variational
Lagrangian density%
\begin{equation}
L_{R}\left( x,\widehat{x},r,s\right) \equiv \pi _{\mu \nu }\frac{D}{Ds}%
g^{\mu \nu }-H_{R}\left( x,\widehat{x},r,s\right) .  \label{legendre-hh}
\end{equation}%
Thus, $L_{R}\left( x,\widehat{x},r,s\right) $ is identified with the
Legendre transform of the corresponding variational Hamiltonian density $%
H_{R}\left( x,\widehat{x},r,s\right) $ defined above, with $\pi _{\mu \nu }%
\frac{D}{Ds}g^{\mu \nu }$ denoting the so-called exchange term. Then the
variational principle associated with the functional $S_{R}\left( x,\widehat{%
x}\right) $\ is prescribed in terms of the synchronous-variation operator $%
\delta $\ (i.e., identified with the Frechet derivative according to Ref.%
\cite{noi1}), i.e., by means of the synchronous variational principle%
\begin{equation}
\delta S_{R}\left( x,\widehat{x}\right) =0
\label{SYNCR-VARIATIONAL-PRINCIPLE}
\end{equation}%
obtained keeping constant both the\ prescribed state $\widehat{x}$ and the $%
4-$scalar volume element $d\Omega .$ This delivers the $4-$tensor
Euler-Lagrange equations cast in symbolic form%
\begin{equation}
\left\{
\begin{array}{c}
\frac{\delta S_{R}\left( x,\widehat{x}\right) }{\delta g^{\mu \nu }}=0, \\
\frac{\delta S_{R}\left( x,\widehat{x}\right) }{\delta \pi _{\mu \nu }}=0,%
\end{array}%
\right.
\end{equation}%
which are manifestly equivalent to the Hamilton equations (\ref{canonical
evolution equations -2}). These equations can be written in the equivalent
Poisson-bracket representation%
\begin{equation}
\frac{D}{Ds}x_{R}(s)=\left[ x_{R},H_{R}\left( x_{R},\widehat{x}%
_{R}(r),r,s\right) \right] _{(x_{R})},
\label{LP-PARAMETRIZED REDUCED HAM-EQ}
\end{equation}%
with $\left[ ,\right] _{(x_{R})}$ denoting the Poisson bracket evaluated
with respect to the canonical variables $x_{R}$, namely%
\begin{equation}
\left[ x_{R},H_{R}\left( x_{R},\widehat{x}_{R}(r),r,s\right) \right]
_{(x_{R})}=\frac{\partial x_{R}}{\partial g^{\mu \nu }}\frac{\partial
H_{R}(s)}{\partial \pi _{\mu \nu }}-\frac{\partial x_{R}}{\partial \pi _{\mu
\nu }}\frac{\partial H_{R}(s)}{\partial g^{\mu \nu }}.
\end{equation}%
Then, after elementary algebra, the PDE's (\ref{LP-PARAMETRIZED REDUCED
HAM-EQ}) yield the GR-Hamilton equations in evolution form given above by
Eqs.(\ref{canonical evolution equations -2}). In particular, invoking Eqs.(%
\ref{KINETIC ENERGY DENSITY})-(\ref{POT-ENERGY-SOURCES-2}) it follows%
\begin{equation}
\frac{\partial V\left( g,\widehat{x}_{R}(r),r,s\right) }{\partial g^{\mu \nu
}(s)}=\sigma \kappa h(s)\widehat{R}_{\mu \nu }-\sigma \kappa g_{\mu \nu }(s)%
\frac{1}{2}\left( g^{\alpha \beta }(s)\widehat{R}_{\alpha \beta }-2\Lambda
\right) -\sigma \kappa \frac{8\pi G}{c^{2}}T_{\mu \nu },  \label{GRADIENT-V}
\end{equation}%
where $\widehat{R}_{\mu \nu }\equiv \widehat{R}_{\mu \nu }(s)$ and $T_{\mu
\nu }\equiv T_{\mu \nu }(s)$ denote the LP-parametrizations of the Ricci and
stress-energy tensors. Hence, in the case the gauge function\ $f(h)$ is
prescribed as $f(h)=1$, the canonical equations (\ref{canonical evolution
equations -2}) reduce to the single equivalent \emph{Lagrangian evolution
equation }for the variational field\emph{\ }$g_{\mu \nu }(s)$ in the
LP-parametrization:%
\begin{equation}
\frac{D}{Ds}\left[ \frac{D}{Ds}g_{\mu \nu }(s)\right] +\sigma h(s)\widehat{R}%
_{\mu \nu }-\sigma g_{\mu \nu }(s)\frac{1}{2}\left[ g^{\alpha \beta }(s)%
\widehat{R}_{\alpha \beta }-2\Lambda \right] -\sigma \frac{8\pi G}{c^{2}}%
T_{\mu \nu }=0.  \label{Lagrangian evolution equation}
\end{equation}%
This concludes the proof that the GR-Hamilton equations (\ref{canonical
evolution equations -2}), as well as the equivalent Lagrangian equation (\ref%
{Lagrangian evolution equation}) are - as expected - both variational.

\subsection{2D - Step \#4: Connection with Einstein field equations}

The connection of the canonical equations (\ref{canonical evolution
equations -2}) with the Einstein theory of GR can\ be obtained under the
assumption that the Hamiltonian density does not depend explicitly on proper
time $s$, \textit{i.e.}, it is actually of the form%
\begin{equation}
H_{R}=H_{R}\left( x_{R},\widehat{x}_{R}(r),r\right) .  \label{AUTONOMY}
\end{equation}%
In this case, one furthermore notices that the identities $\widehat{g}_{\mu
\nu }(s)\widehat{g}^{\mu \nu }(s)=\delta _{\mu }^{\mu }$ and $\frac{D}{Ds}%
\widehat{g}_{\mu \nu }(s)\equiv 0$ hold, so that by construction $\widehat{%
\pi }_{\mu \nu }(s)\equiv 0$ and hence the canonical equation for\ $\widehat{%
\pi }_{\mu \nu }(s)$ (or equivalently Eq.(\ref{Lagrangian evolution equation}%
)) delivers for the prescribed fields%
\begin{equation}
\widehat{R}_{\mu \nu }-\widehat{g}_{\mu \nu }(s)\frac{1}{2}\left[ \widehat{g}%
^{\alpha \beta }(s)\widehat{R}_{\alpha \beta }-2\Lambda \right] =\frac{8\pi G%
}{c^{2}}\widehat{T}_{\mu \nu },  \label{Einstein field equations}
\end{equation}%
which coincides with the Einstein field equations. Therefore, in this
framework the latter are obtained by looking for a stationary solution of
the GR-Hamilton equation (\ref{canonical evolution equations -2}),\textit{\
i.e.}, requiring the initial conditions%
\begin{equation}
\left\{
\begin{array}{c}
g_{\mu \nu }(s_{o})\equiv \widehat{g}_{\mu \nu }(s_{o}), \\
\pi _{\mu \nu }(s_{o})\equiv \widehat{\pi }_{\mu \nu }(s_{o})=0,%
\end{array}%
\right.  \label{INITIAL CONDOITIONS}
\end{equation}%
while requiring furthermore for all $s\in I$%
\begin{equation}
\widehat{\pi }_{\mu \nu }(s)=0.  \label{CONSTRAINT CONDITION}
\end{equation}%
Notice that, in principle, additional extrema may exist for the effective
potential, i.e. such that $\frac{\partial V\left( g,\widehat{x}%
_{R}(r),r,s\right) }{\partial g^{\mu \nu }(s)}=0$. One can show that this
indeed happens, for example, in the case of vacuum, namely letting $\widehat{%
T}_{\mu \nu }\equiv 0.$ Thus, besides $g_{\mu \nu }(s)\equiv \widehat{g}%
_{\mu \nu }$ additional extrema include $g_{\mu \nu }(s)\equiv -\frac{2}{3}%
\widehat{g}_{\mu \nu }$ and the case in which $g_{\mu \nu }(s)$ satisfies
identically the constraint equations $h(s)=0$ and $1-\frac{1}{2}g_{\mu \nu
}(s)\widehat{g}^{\mu \nu }=0$. However, once the initial conditions (\ref%
{INITIAL CONDOITIONS}) are set the stationary solution is unique. The
prerequisite for the existence of such a particular solution is, however,
the validity of the constraint condition (\ref{AUTONOMY}), i.e., the
requirement that the GR-Hamilton equations (\ref{canonical evolution
equations -2}) are autonomous. Such a property is non-trivial. In fact, it
might be in principle violated if non-local effects are taken into account
(see for example Refs.\cite{EPJ2,EPJ8}). Analogous circumstance might arise
due to possible quantum effects. The issue will be further discussed in Part
2.

Finally, for completeness, we mention also the connection between the
reduced Hamiltonian system $\left\{ x_{R},H_{R}\right\} $ defined according
to Eqs.(\ref{REDUCED STATE}) and (\ref{reduced HAMILTONIAN}) and the
representation given in Ref.\cite{noi2} in terms of the "extended"
Hamiltonian system $\left\{ x,H\right\} $ and based on the adoption of the
"extended" canonical state $x\equiv \left\{ g^{\mu \nu },\Pi _{\mu \nu
}^{\alpha }\right\} $. More precisely, the connection is obtained, first, by
the prescription $H=H_{R}$,\ and, second, upon identifying $\Pi _{\mu \nu
}^{\alpha }=t^{\alpha }\pi _{\mu \nu }$. In fact it then follows that $\pi
_{\mu \nu }=t_{\alpha }\Pi _{\mu \nu }^{\alpha }$, so that $\pi _{\mu \nu
}\left( r\right) $ represents the projection of $\Pi _{\mu \nu }^{\alpha
}\left( r\right) $\ along the tangent vector $t_{\alpha }(s)$ to the
background geodesic curve.

\subsection{2E - Step \#5: Alternative Hamiltonian structures}

As indicated above, Eq.(\ref{GRADIENT-V}) together with the GR-Hamilton
equations (\ref{canonical evolution equations -2}) provides the required
connection with the Einstein\ field equations. In practice this means that
any suitably-smooth $4-$scalar function such that%
\begin{equation}
\left. \frac{\partial V\left( g,\widehat{x}_{R}(r),r,s\right) }{\partial
g^{\mu \nu }(s)}\right\vert _{g^{\mu \nu }(s)=\widehat{g}^{\mu \nu
}(s)}=\sigma \kappa \widehat{R}_{\mu \nu }-\sigma \kappa \widehat{g}_{\mu
\nu }(s)\frac{1}{2}\left( \widehat{g}^{\alpha \beta }(s)\widehat{R}_{\alpha
\beta }-2\Lambda \right) -\sigma \kappa \frac{8\pi G}{c^{2}}T_{\mu \nu }=0,
\end{equation}%
realizes an admissible Hamiltonian structure of GR. The choice corresponding
to Eqs.(\ref{KINETIC ENERGY DENSITY}) with the functions $V_{o}\left( g,%
\widehat{x}\right) $ and $V_{F}\left( g,\widehat{x},r\right) $ prescribed
according to Eqs.(\ref{POT-ENERGY-SOURCES-2}) corresponds to the
lowest-order polynomial representation (but still non-linear, and thus
non-trivial) in terms of the variational field $g_{\mu \nu }(s)$ for the
variational Hamiltonian.

However, alternative possible realizations of the Hamiltonian structure $%
\left\{ x_{R},H_{R}\right\} $ can be readily identified. In fact, once the
initial conditions (\ref{INITIAL CONDOITIONS}) are set, alternative possible
realizations of the GR-Hamilton equations (\ref{canonical evolution
equations -2}), leading to the correct realization of the Einstein field
equations, can be achieved. These are obtained introducing a transformation
of the type%
\begin{equation}
\left\{
\begin{array}{c}
g_{\mu \nu }(s)\rightarrow g_{\mu \nu }(s), \\
\pi ^{\mu \nu }(s)\rightarrow \pi ^{\mu \nu }(s)-(s-s_{o})P^{\mu \nu }(%
\widehat{x}_{R}), \\
V\left( g,\widehat{x},r,s\right) \rightarrow V_{1}\left( g,\widehat{x}%
,r,s\right) +U_{o}\left( \widehat{g},\widehat{x},s\right) .%
\end{array}%
\right.  \label{TRANSFORMATIONS}
\end{equation}%
Notice that here the function $U_{o}\left( \widehat{g},\widehat{x},s\right) $
remains in principle arbitrary, so that it can always be determined so that
the extremal value of the potential density is preserved, namely $V\left(
\widehat{g},\widehat{x},r,s\right) =V_{1}\left( \widehat{g},\widehat{x}%
,r,s\right) +U_{o}\left( \widehat{g},\widehat{x},s\right) $. However, $%
\widehat{\pi }^{\mu \nu }(s),P_{\mu \nu }(\widehat{x}_{R})$ and $V_{1}\left(
g,\widehat{x},r,s\right) $ can always be determined so that:

1) the extremal momentum $\widehat{\pi }^{\mu \nu }(s)$ is prescribed so
that
\begin{equation}
\widehat{\pi }^{\mu \nu }(s)=(s-s_{o})P^{\mu \nu }(\widehat{x}_{R});
\label{NON-VANISHING MOMENTUM}
\end{equation}

2) $P^{\mu \nu }(\widehat{x}_{R})$ and $V_{1}\left( g,\widehat{x},r,s\right)
$ are such that%
\begin{eqnarray}
&&\left. \left. \frac{\partial V_{1}\left( g,\widehat{x}_{R}(r),r,s\right) }{%
\partial g^{\mu \nu }(s)}\right\vert _{g^{\mu \nu }(s)=\widehat{g}^{\mu \nu
}(s)}-P_{\mu \nu }(\widehat{x}_{R})=\right.  \notag \\
&&\sigma \kappa \widehat{R}_{\mu \nu }-\sigma \kappa \widehat{g}_{\mu \nu
}(s)\frac{1}{2}\left( \widehat{R}-2\Lambda \right) -\sigma \kappa \frac{8\pi
G}{c^{2}}T_{\mu \nu }.
\end{eqnarray}%
Hence, a particular possible realization which leads to a
functionally-different prescription of the potential density, and hence of
the same Hamiltonian structure, is provided, for example, by the setting%
\begin{equation}
\left\{
\begin{array}{c}
V_{1}\left( g,\widehat{x}_{R}(r),r,s\right) \equiv \kappa hg^{\mu \nu }%
\widehat{R}_{\mu \nu }+V_{F}\left( g,\widehat{x},r\right) , \\
U_{o}\left( \widehat{g},\widehat{x},s\right) =-2\kappa \Lambda , \\
P_{\mu \nu }(\widehat{x}_{R})=-\sigma \kappa \widehat{g}_{\mu \nu
}(s)\Lambda ,%
\end{array}%
\right.
\end{equation}%
with $V_{F}\left( g,\widehat{x},r\right) $ being given by Eq.(\ref%
{POT-ENERGY-SOURCES-2}). The present example means that\ the contribution of
the cosmological constant in the Einstein field equations can also be
interpreted as arising due to a non-vanishing canonical momentum of the form
given by Eq.(\ref{NON-VANISHING MOMENTUM}). Alternatively, a realization of
the Einstein field equations with vanishing cosmological constant ($\Lambda
\equiv 0$) can be achieved in terms of the potential density $V\left( g,%
\widehat{x}_{R}(r),r,s\right) $ of the form given above by Eq.(\ref%
{POT-ENERGY-SOURCES-2}) while setting at the same time%
\begin{equation}
P_{\mu \nu }(\widehat{x}_{R})=\sigma \kappa \widehat{g}_{\mu \nu }(s)\Lambda
,
\end{equation}%
where now $\Lambda $ can be interpreted as an arbitrary real $4-$scalar.

From the previous considerations it follows, however, that if a solution of
the type (\ref{NON-VANISHING MOMENTUM}) is permitted the actual
identification of the variational potential density remains essentially
undetermined. It is however obvious that once the requirement $P_{\mu \nu }(%
\widehat{x}_{R})\equiv 0$, or equivalently the constraint condition
introduced above (\ref{CONSTRAINT CONDITION}) are set, the transformations (%
\ref{TRANSFORMATIONS}) reduce necessarily to the trivial one, namely%
\begin{equation}
\left\{
\begin{array}{c}
g_{\mu \nu }(s)\rightarrow g_{\mu \nu }(s), \\
\pi ^{\mu \nu }(s)\rightarrow \pi ^{\mu \nu }(s), \\
V\left( g,\widehat{x},r,s\right) \rightarrow V_{1}\left( g,\widehat{x}%
,r,s\right) +U_{o}\left( \widehat{g},\widehat{x},s\right) ,%
\end{array}%
\right.
\end{equation}%
which leaves unaffected the CHS. As a consequence, in validity of the
constraint (\ref{CONSTRAINT CONDITION}), the same Hamiltonian structure
remains uniquely determined.

\section{3 - Manifestly-covariant Hamilton-Jacobi theory}

From the results established in the previous section, it follows that,
thanks to the\ realization introduced here for the GR-Hamilton equations of
CCG-theory (i.e. Eqs.(\ref{canonical evolution equations -2})), the same
take the form of dynamical evolution equations.\ This follows as a
consequence of the parametrization in terms of the proper-time $s$ adopted
for all geodetics belonging to the background space-time. This feature
permits one to develop in a standard way, in close analogy with classical
Hamiltonian mechanics, the theory of canonical transformations. Given these
premises, in this section the problem of constructing a Hamilton-Jacobi
theory of GR is addressed. Such a theory should describe a dynamical flow
connecting a generic phase-space state with a suitable initial phase-space
state characterized by identically-vanishing (\textit{i.e., }stationary with
respect to $s$) coordinate fields and momenta. In view of the similarity of
the LP formalism for GR with classical mechanics, it is expected that also
in the present context the Hamilton-Jacobi theory follows from constructing
a symplectic canonical transformation associated with a mixed-variable
generating function of type $S\left( g^{\beta \gamma },P_{\mu \nu },\widehat{%
x}_{R},r,s\right) $.

Accordingly, the transformed canonical state $X_{R}\equiv \left\{ G_{\mu \nu
},P_{\mu \nu }\right\} $ must satisfy the constraint equations%
\begin{eqnarray}
\frac{D}{Ds}P_{\mu \nu }(s_{o}) &=&0, \\
\frac{D}{Ds}G^{\mu \nu }(s_{o}) &=&0,
\end{eqnarray}%
which imply the Hamilton equations%
\begin{eqnarray}
0 &=&\left[ P_{\mu \nu },K_{R}\left( X_{R},\widehat{X}_{R},r,s\right) \right]
_{(X_{R})},  \label{PB-11} \\
0 &=&\left[ G^{\mu \nu },K_{R}\left( X_{R},\widehat{X}_{R},r,s\right) \right]
_{(X_{R})},  \label{PB-127}
\end{eqnarray}%
where $K_{R}\left( X_{R},\widehat{X}_{R},r,s\right) $ is the transformed
Hamiltonian given by%
\begin{equation}
K_{R}\left( X_{R},\widehat{X}_{R},r\right) =H_{R}\left( x_{R},\widehat{x}%
_{R},r,s\right) +\frac{\partial }{\partial s}S\left( g^{\beta \gamma
},P_{\mu \nu },\widehat{x}_{R},r,s\right) .  \label{qqq}
\end{equation}%
Thanks to Eqs.(\ref{PB-11}) and (\ref{PB-127}), the transformed Hamiltonian
is necessarily independent of $X_{R}$. As a consequence, $K_{R}$ identifies
an arbitrary gauge function, \textit{i.e., }in actual fact $%
K_{R}=K_{R}\left( \widehat{x}_{R},r\right) $, which can always be set equal
to zero ($K_{R}=0$). On the other hand, canonical transformation theory
requires that it must be%
\begin{eqnarray}
\pi _{\iota \xi } &=&\frac{\partial S\left( g^{\beta \gamma },P_{\mu \nu },%
\widehat{x}_{R},r,s\right) }{\partial g^{\iota \xi }},  \label{R1-1} \\
G^{\iota \xi } &=&\frac{\partial S\left( g^{\beta \gamma },P_{\mu \nu },%
\widehat{x}_{R},r,s\right) }{\partial P_{\iota \xi }}.  \label{R2-2}
\end{eqnarray}%
Then, introducing the $s-$parametrization it follows that Eq.(\ref{qqq})
delivers%
\begin{equation}
H_{R}\left( g^{\beta \gamma },\frac{\partial S\left( g^{\beta \gamma
},P_{\mu \nu },\widehat{x}_{R},s\right) }{\partial g^{\iota \xi }},\widehat{x%
}_{R},r,s\right) +\frac{\partial }{\partial s}S\left( g^{\beta \gamma
},P_{\mu \nu },\widehat{x}_{R},r,s\right) =0,  \label{AA-REDUCED-HJ}
\end{equation}%
which realizes the GR-Hamilton-Jacobi equation for the mixed-variable
generating function $S\left( g^{\beta \gamma },P_{\mu \nu },\widehat{x}%
_{R},r,s\right) $. Due to its similarity with the customary Hamilton-Jacobi
equation well-known in Hamiltonian classical dynamics, in the following $S$
will be referred to as the (\emph{classical}) \emph{GR-Hamilton principal
function}. The canonical transformations generated by $S\left( g^{\beta
\gamma },P_{\mu \nu },\widehat{x}_{R},s\right) $ are then obtained by the
set of equations (\ref{R1-1}), (\ref{R2-2}) and (\ref{AA-REDUCED-HJ}).

Now we notice, in view of the discussions given above, that the inverse
canonical transformation $X_{R}\rightarrow x_{R}$ locally exists provided
the invertibility condition on the Hessian determinant $\det \left\vert %
\left[ \frac{\partial ^{2}S\left( g^{\beta \gamma },P_{\mu \nu },\widehat{x}%
_{R},r,s\right) }{\partial g^{\rho \sigma }\partial P_{\iota \xi }}\right]
_{X_{R}=\widehat{x}_{R}}\right\vert \neq 0$ is met. Under such a condition
the direct canonical equation (\ref{R2-2}) determines $g^{\beta \gamma }$ as
an implicit function of the form $g^{\beta \gamma }=g^{\beta \gamma }\left(
G^{\beta \gamma },P_{\mu \nu },\widehat{x}_{R},r,s\right) $.

The following statement holds on the relationship between the
GR-Hamilton-Jacobi and the GR-Hamilton equations.

\bigskip

\textbf{THM.1 - Equivalence of GR-Hamilton and GR-Hamilton-Jacobi equations}

\emph{The GR-Hamilton-Jacobi equation (\ref{AA-REDUCED-HJ}) subject to the
constraint (\ref{R1-1}) is equivalent to the set of GR-Hamilton equations
expressed in terms of the initial canonical variables, as given by Eqs.(\ref%
{canonical evolution equations -2}).}

\emph{Proof - }Without loss of generality and avoiding possible
misunderstandings, the compact notation $S\left( g,P,\widehat{x}%
_{R},r,s\right) $ will be used in the following proof to denote the
GR-Hamilton principal function. To start with, we evaluate first the partial
derivative of Eq.(\ref{AA-REDUCED-HJ}) with respect to $g^{ik}$, keeping
both $\frac{\partial S\left( g,P,\widehat{x}_{R},r,s\right) }{\partial
g^{\iota \xi }}$ and $r^{\mu }$ constant. This gives%
\begin{equation}
\frac{\partial }{\partial g^{ik}}H_{R}\left( g^{\beta \gamma },\frac{%
\partial S\left( g,P,\widehat{x}_{R},s\right) }{\partial g^{\iota \xi }},%
\widehat{x}_{R},r,s\right) +\frac{\partial }{\partial s}\left[ \frac{%
\partial }{\partial g^{ik}}S\left( g,P,\widehat{x}_{R},r,s\right) \right]
_{\left( g,P\right) }=0.  \label{first-hjh}
\end{equation}%
Then, let us evaluate in a similar manner the partial derivative with
respect to $\frac{\partial S\left( g,P,\widehat{x}_{R},s\right) }{\partial
g^{ik}}$, keeping $g^{\mu \nu }$ and $r^{\mu }$ constant. This gives%
\begin{equation}
\frac{\partial }{\partial \frac{\partial S\left( g,P,\widehat{x}%
_{R},r,s\right) }{\partial g^{ik}}}H_{R}\left( g^{\beta \gamma },\frac{%
\partial S\left( g,P,\widehat{x}_{R},s\right) }{\partial g^{\iota \xi }},%
\widehat{x}_{R},r,s\right) +\left[ \frac{\partial }{\partial \frac{\partial
S\left( g,P,\widehat{x}_{R},r,s\right) }{\partial g^{ik}}}\frac{\partial }{%
\partial s}S\left( g,P,\widehat{x}_{R},r,s\right) \right] _{\left(
g,P\right) }=0.  \label{second-hjh}
\end{equation}%
With the identification $\pi _{\iota \xi }=\frac{\partial S\left( g,P,%
\widehat{x}_{R},s\right) }{\partial g^{\iota \xi }}$ provided by Eq.(\ref%
{R1-1}) it follows that Eq.(\ref{first-hjh}) becomes%
\begin{equation}
\frac{\partial }{\partial g^{ik}}H_{R}\left( g^{\beta \gamma },\pi _{\iota
\xi },\widehat{x}_{R},r\right) +\frac{D}{Ds}\pi _{ik}=0,  \label{HAM-EQ-2a}
\end{equation}%
which coincides with the second Hamilton equation in (\ref{canonical
evolution equations -2}). To prove also the validity of the Hamilton
equation for $g^{\beta \gamma }$ we first invoke the following identity%
\begin{gather}
\left[ \frac{\partial }{\partial \frac{\partial S\left( g,P,\widehat{x}%
_{R},s\right) }{\partial g^{ik}}}\frac{\partial }{\partial s}S\left( g,P,%
\widehat{x}_{R},s\right) \right] _{\left( g,P\right) }=\frac{\partial }{%
\partial \frac{\partial S\left( g,P,\widehat{x}_{R},s\right) }{\partial
g^{ik}}}\frac{\partial }{\partial s}S\left( g,P,\widehat{x}_{R},s\right)
\notag \\
-\frac{D}{Ds}g^{\beta \gamma }\frac{\partial }{\partial \frac{\partial
S\left( g,P,\widehat{x}_{R},s\right) }{\partial g^{ik}}}\frac{\partial
S\left( g,P,\widehat{x}_{R},s\right) }{\partial g^{\beta \gamma }},
\label{ts}
\end{gather}%
where%
\begin{equation}
\frac{\partial }{\partial \frac{\partial S\left( g,P,\widehat{x}%
_{R},r,s\right) }{\partial g^{ik}}}\frac{\partial S\left( g,P,\widehat{x}%
_{R},r,s\right) }{\partial g^{\beta \gamma }}=\delta _{\beta }^{i}\delta
_{\gamma }^{k}.
\end{equation}%
The first term on the rhs of Eq.(\ref{ts}) vanishes identically because $%
\frac{\partial }{\partial s}S\left( g,P,\widehat{x}_{R},r,s\right) $ must be
considered as independent of $\pi _{ik}$. Therefore, Eq.(\ref{second-hjh})
gives%
\begin{equation}
\frac{\partial }{\partial \pi _{ik}}H_{R}\left( g^{\beta \gamma },\pi
_{\iota \xi },\widehat{x}_{R},r,s\right) -\frac{D}{Ds}g^{ik}=0,
\label{HAM-EQ-1}
\end{equation}%
which coincides with the Hamilton equation for $g^{ik}$ and gives also the
relationship of the generalized velocity $\frac{D}{Ds}g^{ik}$ with the
canonical momentum, since here no explicit $s-$dependence appears. This
proves the equivalence between the GR-Hamilton-Jacobi and GR-Hamilton
equations, both expressed in manifestly-covariant form. \textbf{Q.E.D.}

\bigskip

This conclusion recovers the relationship between Hamilton and
Hamilton-Jacobi equations holding in Hamiltonian Classical Mechanics for
discrete dynamical systems. The connection is established also in the
present case for the continuum gravitational field thanks to the
manifestly-covariant LP parametrization of the theory and the representation
of the Hamiltonian and Hamilton-Jacobi equations as dynamical evolution
equations with respect to the proper time $s$ characterizing background
geodetics. The physical interpretation which follows from the validity of
THM.1 is remarkable. This concerns the meaning of the Hamilton-Jacobi theory
in providing a wave mechanics description of the continuum Hamiltonian
dynamics. This follows also in the present context by comparing the
mathematical structure of the Hamilton-Jacobi equation (\ref{AA-REDUCED-HJ})
with the well-known eikonal equation of geometrical optics. In fact Eq.(\ref%
{AA-REDUCED-HJ}) contains the squared of the derivative $\frac{\partial
S\left( g^{\beta \gamma },P_{\mu \nu },\widehat{x}_{R},r,s\right) }{\partial
g^{\iota \xi }}$, so that the Hamilton principal function $S\left( g^{\beta
\gamma },P_{\mu \nu },\widehat{x}_{R},r,s\right) $ is associated with the
eikonal (\textit{i.e., }the phase of the wave), while the remaining
contributions due to the geometrical and physical properties of the curved
space-time formally play the role of a non-uniform index of refraction in
geometrical optics \cite{Goldstein}. The outcome pointed out here proves
that the dynamics of the field $g_{\mu \nu }(s)$ in the virtual domain of
variational fields where the Hamiltonian structure is defined and the
Hamilton-Jacobi theory (\ref{AA-REDUCED-HJ}) applies must be characterized
by a wave-like behavior and can therefore be given a geometrical optics
interpretation. This feature is expected to be crucial for the establishment
of the corresponding manifestly-covariant quantum theory of the
gravitational field.

An important\ qualitative feature must be pointed out regarding the
Hamilton-Jacobi theory developed here. This\ refers to a formal difference
arising between the Hamilton-Jacobi theory for continuum fields built on the
DeDonder-Weyl covariant approach and the Hamilton-Jacobi theory holding in
classical mechanics for particle dynamics. This concerns, more precisely,
the dimensional units to be adopted for the Hamilton principal function $S$
and hence also the canonical momentum $\pi ^{\mu \nu }$. Indeed, as is
well-known, in particle\textbf{\ }dynamics $S$\ retains the dimension of an
action (and therefore of the action functional), so that $\left[ S\right] =%
\left[ \hbar \right] $. In the present case instead (see Eq.(\ref{reduced
HAMILTONIAN})) one has that $\left[ S\right] =\left[ \hbar L^{-3}\right] $,
namely the dimension of $S$\ differs from that of an action by the cubic
length $L^{-3}$. This arises because for continuum fields the action
functional is an integral over the $4-$volume element of the Hamiltonian
density, while for particle mechanics it is expressed as a line integral
over the proper-time length. One has to notice, however, that, first, the
dimensions of $S$\ may be changed by the introduction of a non-symplectic
canonical transformation. This means that, by suitable choice of the same
transformation, $S$ can actually recover the dimension of an action. Second,
the relationship between the Hamilton principal function $S$\ and the
Hamiltonian function itself remains in all cases the same, with the two
functions differing by the dimension of a length (see also Section 4 below).

Before concluding, the following additional remarks are in order:

1) The GR-Hamilton-Jacobi description permits to construct explicitly
canonical transformations mapping in each other the physical and virtual
domains. The generating function determined by the GR-Hamilton-Jacobi
equation is a real $4-$scalar field.

2)\ The generating function obtained in this way realizes the particular
subset of canonical transformations which map the physically-observable
state $\widehat{x}_{R}$\ into a neighboring admissible virtual canonical
state $x_{R}$.

3) The virtue of the approach is that it preserves the validity of the
Einstein equation in the physical domain. In other words, the canonical
transformations do not affect the physical behavior.

4) A further issue concerns the connection between the same prescribed
metric tensor $\widehat{g}(r)\equiv \left\{ \widehat{g}_{\mu \nu
}(r)\right\} $\ and the variational/extremal state $x_{R}(s)=\left\{
g(s),\pi (s)\right\} $. This can be obtained by establishing a proper
statistical theory achieved by considering the initial state\textbf{\ }%
\begin{equation}
x_{R}(s_{o})=\widehat{x}(s_{o})+\delta x_{R}(s_{o})
\end{equation}%
as a stochastic tensor and thus endowed with a suitable phase-space
probability density. The topic can be developed in the framework of a
statistical description of classical gravity to be discussed elsewhere in
detail.

\section{4 - Properties of CHS}

In this section some properties are discussed which characterize the
manifestly-covariant Hamiltonian theory of GR.

\subsection{4A - Global prescription and regularity}

This refers, first of all, to the global prescription and regularity of the
GR-Lagrangian and GR-Hamiltonian densities $L_{R}\equiv L_{R}(y,\widehat{g}%
,r,s)$ and $H_{R}\equiv H_{R}(x,\widehat{g},r,s)$ defined according to Eqs.(%
\ref{legendre-hh}) and (\ref{reduced HAMILTONIAN})\ which are associated
with the corresponding Hamiltonian structure $\left\{ x_{R},H_{R}\right\} $\
indicated above. For this purpose we notice that in Eq.(\ref{legendre-hh})
the effective kinetic and potential densities expressed in terms of the
Lagrangian state $y=\left\{ g_{\mu \nu },\frac{D}{Ds}g^{\mu \nu }\right\} $
and the LP-parametrization (\ref{LP-PARAMETRIZED REDUCED HAM-EQ}) can be
taken of the general type%
\begin{equation}
\left\{
\begin{array}{c}
T_{R}\left( y,\widehat{g}\right) =\frac{f(h)}{2\kappa }\frac{D}{Ds}g_{\mu
\nu }\frac{D}{Ds}g^{\mu \nu }, \\
V\left( g,\widehat{g},r,s\right) \equiv \sigma V_{o}\left( g,\widehat{g}%
\right) +\sigma V_{F}\left( g,\widehat{g},r,s\right) .%
\end{array}%
\right.  \label{kin-energy-density}
\end{equation}%
Here, $f(h)$ and $\sigma $ identify the two distinct multiplicative gauge
functions\ introduced above (see Section 2), $\kappa $ is the dimensional
constant $\kappa =\frac{c^{3}}{16\pi G}$, while the rest of the notations is
expressed in standard form according to Refs.\cite{noi1,noi2}. More
precisely, in the second equation $V_{o}\left( g,\widehat{g}\right) $ and $%
V_{F}\left( g,\widehat{g},r,s\right) \equiv hL_{F}\left( g,\widehat{g}%
,r,s\right) $ are defined as in Eq.(\ref{POT-ENERGY-SOURCES-2}) and must be
expressed here in Lagrangian variables, with the field Lagrangian $%
L_{F}\left( g,\widehat{g},r,s\right) $ being prescribed according to Ref.%
\cite{noi1}. Furthermore,\textbf{\ }$T_{R}\left( y,\widehat{g}\right) $
identifies the \emph{generic form} of the effective kinetic density. It
follows that a sufficient condition for the global prescription of the
canonical state, \textit{i.e., }the existence of a smooth bijection
connecting the Lagrangian and Hamiltonian states is the so-called \emph{%
regularity condition} of the GR-Hamiltonian (and corresponding
GR-Lagrangian) density. This requires more precisely that in the whole
Hamiltonian phase-space%
\begin{equation}
\left\vert \frac{\partial ^{2}H_{R}}{\partial \pi _{\mu \nu }\partial \pi
^{\alpha \beta }}\right\vert \equiv \left\vert \frac{\partial ^{2}T_{R}}{%
\partial \pi _{\mu \nu }\partial \pi ^{\alpha \beta }}\right\vert =\frac{1}{%
\kappa f(h)}\neq 0.  \label{Hessian-1}
\end{equation}

\subsection{4B - Gauge indeterminacies of CHS}

As shown in Ref.\cite{noi2} at the classical level\ the Hamiltonian
structure $\left\{ x_{R},H_{R}\right\} $ of SF-GR remains intrinsically
non-unique, with the Hamiltonian density $H_{R}$ being characterized by
suitable gauge indeterminacies. Leaving aside the treatment of additive
gauge functions earlier discussed in Refs.\cite{noi1,noi2}, these refer\
more precisely to the following properties:

\begin{itemize}
\item \emph{A) The first one is the so-called multiplicative gauge
transformation of the effective kinetic density.} To identify it we notice
that the scalar factor $f(h)$ appearing in the prescriptions of the
effective kinetic density (see first equation in (\ref{kin-energy-density}))
remains in principle essentially indeterminate. In fact the regularity
condition (\ref{Hessian-1}) requires only that
\begin{equation}
\left\vert \frac{\partial ^{2}T_{R}}{\partial \pi _{\mu \nu }\partial \pi
^{\alpha \beta }}\right\vert =\frac{1}{\kappa f(h)}\neq 0,
\end{equation}%
implying that the function $f(h)$ can be realized by an arbitrary
non-vanishing (for example, strictly positive) and suitably smooth
dimensionless real function. In addition, in order that both $T_{R}\left( y,%
\widehat{g}\right) $ and $V\left( y,\widehat{g}\right) $ (see again Eqs.(\ref%
{kin-energy-density}))\textbf{\ }are realized by means of integrable
functions in the configurations space $U_{g}$, accordingly the functions $%
f(h)$ and $1/f(h)$ should be summable too. As a consequence the prescription
of $f(h)$ remains in principle still free within the classical theory of
SF-GR developed in Section 2. This means that $f(h)$ should be intended in
such a context as a gauge indeterminacy affecting the Hamiltonian density $%
H_{R}$, \textit{i.e., }with respect to the \emph{multiplicative gauge
transformation}%
\begin{equation}
\left\{
\begin{array}{c}
T_{R}\left( x_{R},\widehat{g}\right) \equiv \frac{1}{2\kappa }\pi _{\mu \nu
}\pi ^{\mu \nu }\rightarrow T_{R}^{\prime }\left( x_{R},\widehat{g}\right)
\equiv \frac{1}{2\kappa f(h)}\pi _{\mu \nu }^{\prime }\pi ^{\prime \mu \nu }=%
\frac{1}{f(h)}T_{R}\left( x_{R},\widehat{g}\right) , \\
\pi _{\mu \nu }\rightarrow \pi _{\mu \nu }^{\prime }=f(h)\pi _{\mu \nu }.%
\end{array}%
\right.  \label{multiplicative-gauge-1}
\end{equation}

\item \emph{B) The second indeterminacy is related to the so-called
multiplicative gauge transformation of the effective potential density }$%
V\left( g,\widehat{g},r,s\right) $ (see again Section 2). More precisely the
indeterminacy is related to the constant gauge factor $\sigma =\pm 1$.

\item \emph{C) The third indeterminacy is related to the so-called additive
gauge transformation.} Indeed, one can readily show (see Ref.\cite{noi2})
that $L_{R}(y,\widehat{g},s)$ is prescribed up to an arbitrary additive
gauge transformation of the type%
\begin{equation}
L_{R}(y,\widehat{g},r,s)\rightarrow L_{R}(y,\widehat{g},r,s)+\Delta V
\label{classical-gauge-2}
\end{equation}%
being $\Delta V$ a gauge scalar field of the form $\Delta V=\frac{DF(g,%
\widehat{g},r,s)}{Ds}$, with $F(g,\widehat{g},s)$ being an arbitrary,
suitably-smooth real gauge function of class $C^{(2)}$ with respect to the
variables $(g,s)$ (see also related discussion in Ref.\cite{noi1}).
\end{itemize}

\subsection{4C - Dimensional normalization of CHS}

In this section it is shown that the Hamiltonian structure\textbf{\ }$%
\left\{ x_{R},H_{R}\right\} $ can be equivalently realized in such a way
that $x_{R}$, and consequently also $H_{R}$, can be suitably normalized,
\textit{i.e., }so that to achieve prescribed physical dimensions. Granted
the non-symplectic canonical nature of the transformation indicated above (%
\textit{i.e., }Eq.(\ref{CANONICAL-0})) one can always identify the $4-$%
scalar $\alpha $ with a classical invariant parameter, \textit{i.e., }both
frame-independent and space-time independent. In particular, in the
framework of a classical treatment it should be identified with the
classical parameter $\alpha \equiv \alpha _{\text{Classical}},$ being
\begin{equation}
\alpha _{\text{Classical}}=m_{o}cL>0,  \label{ALFA-1}
\end{equation}%
with $c$ being the speed of light in vacuum, $m_{o}$ a suitable rest-mass
(to be later identified with the non-vanishing graviton mass in the
framework of quantum theory of GR) and $L$\ a characteristic scale length to
be considered as an invariant non-null $4-$scalar.\ Without loss of
generality this can always be assumed of the form\textbf{\ }%
\begin{equation}
L=L(m_{o}),  \label{charact-lewngth}
\end{equation}%
with $m_{o}$ itself being regarded as an invariant $4-$scalar. Here $L$\ is
regarded as a classical invariant parameter, so that it should remain
independent of all quantum parameters, \textit{i.e., }in particular $\hslash
$.\ In addition, in view of the covariance property of the theory, whereby
the choice of the background space-time $\left( \mathbf{Q}^{4},\widehat{g}%
(r)\right) $ is in principle\ arbitrary, the $4-$scalars $m_{o}$\ and $L$\
should be \emph{universal constants,}\textbf{\ }namely also invariant with
respect to the action of local and non-local point transformations \cite%
{noi4}. As an example, a possible consistent choice for the invariant
function $L(m_{o})$\ is realized by means of the so-called Schwarzschild
radius, \textit{i.e.,}%
\begin{equation}
L(m_{o})=\frac{2m_{o}G}{c^{2}}.  \label{Schwartzchild radius}
\end{equation}%
The invariant rest-mass $m_{o}$\ remains however still arbitrary at this
level, its prescription being left to quantum theory. It follows that the
transformed GR-Hamilton equations (\ref{canonical evolution equations -2})
can always be cast in the\emph{\ dimensional normalized form}%
\begin{equation}
\left\{
\begin{array}{c}
\frac{Dg_{\mu \nu }}{Ds}=\frac{\partial \overline{H}_{R}}{\partial \overline{%
\pi }^{\mu \nu }}\equiv \frac{\partial H_{R}}{\partial \pi ^{\mu \nu }}, \\
\frac{D\overline{\pi }_{\mu \nu }}{Ds}=-\frac{\partial \overline{H}_{R}}{%
\partial g^{\mu \nu }}\equiv -\frac{\alpha L}{k}\frac{\partial H_{R}}{%
\partial g^{\mu \nu }},%
\end{array}%
\right.  \label{can-2-a}
\end{equation}%
where the transformed Hamiltonian $\overline{H}_{R}$ identifies the \emph{%
normalized GR-Hamiltonian density}%
\begin{equation}
\overline{H}_{R}(\overline{x}_{R},\widehat{g},r,s)=\frac{1}{f(h)}\overline{T}%
_{R}(\overline{x}_{R},\widehat{g},r,s)+\overline{V}\left( \overline{g},%
\widehat{g},r,s\right) .  \label{HAmiltonian-N}
\end{equation}%
Here the notation is as follows. First, for an arbitrary curved space-time $(%
\mathbf{Q}^{4},\widehat{g}(r))$ the functions $\overline{T}_{R}$ and $%
\overline{V}$ now are identified with%
\begin{equation}
\left\{
\begin{array}{c}
\overline{T}_{R}(\overline{x}_{R},\widehat{g},r,s)=\frac{\overline{\pi }%
^{\mu \nu }\overline{\pi }_{\mu \nu }}{2\alpha L}, \\
\overline{V}\left( g,\widehat{g},r,s\right) \equiv \sigma \overline{V}%
_{o}\left( g,\widehat{g},r,s\right) +\sigma \overline{V}_{F}\left( g,%
\widehat{g},r,s\right) ,%
\end{array}%
\right.  \label{V_R-N}
\end{equation}%
so that $\overline{T}_{R}(\overline{x}_{R},\widehat{g},r,s)$ identifies the
normalized effective kinetic density and $\overline{V}$ by analogy is the
corresponding normalized effective potential density, with $\overline{V}%
_{o}\left( g,\widehat{g}\right) $ and $\overline{V}_{F}\left( g,\widehat{g}%
,r,s\right) $ now being prescribed respectively in terms of $V_{o}\left( g,%
\widehat{g}\right) $ and $V_{F}\left( g,\widehat{g},r,s\right) $ as
\begin{equation}
\left\{
\begin{array}{c}
\overline{V}_{o}\left( g,\widehat{g}\right) \equiv h\alpha L\left[ g^{\mu
\nu }\widehat{R}_{\mu \nu }-2\Lambda \right] , \\
V_{F}\left( g,\widehat{g},r,s\right) \equiv \frac{h\alpha L}{2k}L_{F}\left(
g,\widehat{g},r,s\right) .%
\end{array}%
\right.  \label{V_F-N}
\end{equation}%
From the canonical equations (\ref{can-2-a}) it is obvious that by
construction the transformed canonical momentum $\overline{\pi }^{\mu \nu }$
takes the dimensions of an action, \textit{i.e., }$\left[ \overline{\pi }%
^{\mu \nu }\right] =\left[ \hslash \right] $. The set $\left\{ \overline{x}%
_{R},\overline{H}_{R}\right\} $ thus provides an admissible representation
of CHS. In particular it follows that the GR-Hamilton equations in
normalized form become respectively%
\begin{equation}
\left\{
\begin{array}{c}
\frac{Dg_{\mu \nu }}{Ds}=\frac{\overline{\pi }_{\mu \nu }}{\alpha L}, \\
\frac{D\overline{\pi }_{\mu \nu }}{Ds}=-\frac{\partial \overline{V}\left( g,%
\widehat{g},r,s\right) }{\partial g^{\mu \nu }},%
\end{array}%
\right.  \label{RENORM-HAM;}
\end{equation}%
with the operator $D/Ds$\ being prescribed again in terms of the
corresponding prescribed metric tensors $\widehat{g}_{\mu \nu }(r)$.

For completeness, we mention here also the normalized Hamilton-Jacobi
equation corresponding to the canonical equations (\ref{RENORM-HAM;}). This
is reached introducing the corresponding normalized Hamilton principal
function $S(g,P,\widehat{g},r,s)$, \textit{i.e., }the mixed-variable
generating function for the canonical transformation%
\begin{equation}
x_{R}(s_{o})\equiv (\mathbf{G}_{\mu \nu },\mathbf{P}^{\mu \nu
})\Leftrightarrow x(s)\equiv (g_{\mu \nu }(s),\pi ^{\mu \nu }(s)),
\end{equation}%
with $x_{R}(s_{o})\equiv (\mathbf{G}_{\mu \nu },\mathbf{P}^{\mu \nu })$
denoting the initial canonical GR-state. Then, $S(g,\mathbf{P},\widehat{g}%
,r,s)$ is prescribed in such a way that the normalized canonical momentum $%
\pi ^{\mu \nu }(s)$ is given by $\overline{\pi }_{\mu \nu }=\frac{\partial
S(g,\mathbf{P},\widehat{g},r,s)}{\partial g^{\mu \nu }}$, while the initial
canonical coordinate $\mathbf{G}_{\mu \nu }$ is determined by the inverse
canonical transformation $\mathbf{G}_{\mu \nu }=\frac{\partial S(g,\mathbf{P}%
,\widehat{g},r,s)}{\partial \mathbf{P}^{\mu \nu }}$. It follows that the
corresponding dimensionally-normalized Hamilton-Jacobi equation which is
equivalent to Eqs.(\ref{RENORM-HAM;}) is provided by%
\begin{equation}
\frac{\partial S(g,\mathbf{P},\widehat{g},r,s)}{\partial s}+\overline{H}%
_{R}\left( g,\overline{\pi }\equiv \frac{\partial S(g,\mathbf{P},\widehat{g}%
,r,s)}{\partial g},\widehat{g},r,s\right) =0,  \label{GR-HAM-JACOBI}
\end{equation}%
with $\overline{H}_{R}$ being prescribed by Eqs.(\ref{HAmiltonian-N}).

\section{5 - Structural stability of the GR-Hamilton equations}

In this section we present an application of the GR-Hamiltonian theory for
the Einstein field equations developed in this paper, which is represented
by Eqs.(\ref{canonical evolution equations -2}) or equivalently by the
Hamilton-Jacobi equation (\ref{AA-REDUCED-HJ}). This refers to the stability
of the GR-Hamilton equations with respect to their stationary solution. As
shown above the latter realizes by construction a solution of the Einstein
field equations (\ref{Einstein field equations})\ in its most general form,
i.e., in the presence of arbitrary external sources. Therefore the task to
be addressed concerns the so-called \emph{structural stability} of the
GR-Hamilton equations (with respect to the Einstein field equations), namely
the stability of stationary solutions of the GR-Hamilton equations assuming
that the perturbed fields realize particular solutions of the same
GR-Hamilton equations.

As a first illustration of the problem, here we consider the case of
arbitrary vacuum solutions realized by setting a vanishing stress-energy
tensor ($\widehat{T}_{\mu \nu }=0$)\ and possibly retaining also a
non-vanishing cosmological constant as corresponds to de Sitter ($\Lambda >0$%
) or anti-de Sitter ($\Lambda <0$) space-times.

Let us address the problem in the context of the reduced continuum
manifestly-covariant Hamiltonian theory. The study is supported by the
conclusions concerning the wave mechanics interpretation of the reduced
continuum Hamiltonian dynamics of the gravitational field determined by the
Hamilton-Jacobi theory. For this purpose we shall consider perturbations of
the reduced canonical state $x_{R}(s)$ which are suitably close to $\widehat{%
x}_{R}(s)$, namely of the form%
\begin{eqnarray}
g^{\mu \nu }(s) &=&\widehat{g}^{\mu \nu }(s)+\varepsilon \delta g^{\mu \nu
}(s),  \label{SOL-1} \\
\pi _{\mu \nu }(s) &=&\varepsilon \delta \pi _{\mu \nu }(s),  \label{SOL-2}
\end{eqnarray}%
where $\varepsilon \ll 1$ is an infinitesimal dimensionless parameter
identifying the perturbations of the canonical fields. In particular,
consistent with the existence of a Hamilton-Jacobi theory and its physical
interpretation pointed out above, we are authorized to consider a wave-like
form of the perturbations. These are assumed to propagate along field
geodetics, namely the same perturbations of the canonical fields $\left(
\delta g^{\mu \nu }(s),\delta \pi _{\mu \nu }(s)\right) $ are taken of the
form%
\begin{eqnarray}
\delta g^{\mu \nu }(s) &=&\delta \widehat{g}^{\mu \nu }(\widehat{g}(s))\exp
\left\{ G(s)\right\} ,  \label{PERT-1} \\
\delta \pi _{\mu \nu }(s) &=&\delta \widehat{\pi }_{\mu \nu }(\widehat{g}%
(s))\exp \left\{ G(s)\right\} .  \label{PERT-2}
\end{eqnarray}%
Here $G(s)$ denotes the eikonal%
\begin{equation}
G(s)=i\frac{\omega }{c}s-iKr^{\mu }(s)t_{\mu }(s),  \label{eiko}
\end{equation}%
with $\omega $\ and $K$ being $4-$scalar parameters which, by construction,
have respectively the dimensions of a frequency and that of a wave number (%
\textit{i.e., }the inverse of a length). Therefore, denoting $K\equiv
1/\lambda $, according to the representation (\ref{eiko}), $\omega $ and $%
\lambda $ identify the invariant frequency and wave-length of the wave-like
perturbations of the canonical fields. The invariant character of $\omega $
and $\lambda $ is a characteristic feature of the manifestly-covariant
Hamiltonian theory.

It is then immediate to show that, in terms of the canonical evolution
equations (\ref{LP-PARAMETRIZED REDUCED HAM-EQ}), the following set of
linear differential equations advancing in proper time the perturbed fields $%
\delta g^{\mu \nu }(s)$ and $\delta \pi ^{\mu \nu }(s)$ and accurate through
$O\left( \varepsilon \right) $ is obtained\ thanks also to the requirement (%
\ref{COINSTRAIUNT ON f(h)}):%
\begin{eqnarray}
\frac{D}{Ds}\delta g^{\mu \nu }(s) &=&\frac{1}{\kappa }\delta \pi ^{\mu \nu
}(s),  \label{prima} \\
\frac{D}{Ds}\delta \pi _{\mu \nu }\left( s\right) &=&\frac{\sigma }{2}\kappa
\left( \widehat{R}+2\Lambda \right) \delta g_{\mu \nu }\left( s\right) +%
\frac{\sigma }{2}\kappa \left( \widehat{g}_{\mu \nu }(s)\widehat{R}_{\alpha
\beta }+\widehat{R}_{\mu \nu }\widehat{g}_{\alpha \beta }(s)\right) \delta
g^{\alpha \beta }.  \label{seconda}
\end{eqnarray}%
In particular, introducing the representation (\ref{eiko}) and recalling the
definition of the differential operator $\frac{D}{Ds}$, the first equation
yields a unique relationship between $\delta \pi ^{\mu \nu }(s)$ and $\delta
g^{\mu \nu }(s),$ namely $\delta \pi ^{\mu \nu }(s)=i\kappa \left[ \frac{%
\omega }{c}-K\right] \delta g^{\mu \nu }(s).$ Then, Eq.(\ref{seconda})
determines the algebraic linear equation for $\delta g_{\mu \nu }\left(
s\right) $:%
\begin{equation}
\left( -\kappa \left[ \frac{\omega }{c}-K\right] ^{2}-\frac{\sigma }{2}%
\kappa \left[ \widehat{R}+2\Lambda \right] \right) \delta g_{\mu \nu }(s)=%
\frac{\sigma \kappa }{2}\left( \widehat{g}_{\mu \nu }(s)\widehat{R}_{\alpha
\beta }+\widehat{R}_{\mu \nu }\widehat{g}_{\alpha \beta }(s)\right) \delta
g^{\alpha \beta }\left( s\right) .
\label{single equation for the perturbation}
\end{equation}

To solve it explicitly one needs to determine the corresponding algebraic
equations holding for the independent tensor products $\widehat{g}_{\mu \nu
}(s)\delta g^{\alpha \beta }\widehat{R}_{\alpha \beta }$ and $\widehat{g}%
_{\alpha \beta }(s)\delta g^{\alpha \beta }\widehat{R}_{\mu \nu }$ appearing
on the rhs of the same equation. Thus, by first multiplying tensorially term
by term Eq.(\ref{single equation for the perturbation}) by $\widehat{R}^{\mu
\nu }$it follows%
\begin{equation}
\left( -\left[ \frac{\omega }{c}-K\right] ^{2}-\sigma \left[ \widehat{R}%
+\Lambda \right] \right) \widehat{R}^{\mu \nu }\delta g_{\mu \nu }\left(
s\right) =\frac{\sigma }{2}\widehat{R}_{\mu \nu }\widehat{R}^{\mu \nu }%
\widehat{g}_{\alpha \beta }(s)\delta g^{\alpha \beta }\left( s\right) .
\label{second single equationm}
\end{equation}%
Next, multiplying tensorially term by term Eq.(\ref{single equation for the
perturbation}) by $\widehat{g}^{\mu \nu }\left( s\right) $ one obtains
instead%
\begin{equation}
\left( -\left[ \frac{\omega }{c}-K\right] ^{2}-\sigma \left[ \widehat{R}%
+\Lambda \right] \right) \widehat{g}^{\mu \nu }\left( s\right) \delta g_{\mu
\nu }\left( s\right) =2\sigma \widehat{R}_{\alpha \beta }\delta g^{\alpha
\beta }\left( s\right) .  \label{third single equation}
\end{equation}%
Combining together equations (\ref{second single equationm}) and (\ref{third
single equation}) one is finally left with the equation
\begin{equation}
\left( -\left[ \frac{\omega }{c}-K\right] ^{2}-\sigma \left[ \widehat{R}%
+\Lambda \right] \right) ^{2}-\widehat{R}_{\mu \nu }\widehat{R}^{\mu \nu }=0,
\label{DISPERSION RELATION}
\end{equation}%
which identifies the dispersion relation between $\omega $ and $K$, \textit{%
i.e., }the condition under which in the context for the reduced continuum
Hamiltonian theory the infinitesimal perturbations (\ref{PERT-1}) and (\ref%
{PERT-2}) can occur.

To analyze in terms of Eq.(\ref{DISPERSION RELATION}) the conditions for the
existence of stable, marginally stable or unstable oscillatory solutions for
the canonical fields $\left( \delta g^{\mu \nu }(s),\delta \pi _{\mu \nu
}(s)\right) $, it is convenient to classify the possible complex roots for $%
\omega $ which can locally occur. Such a classification has therefore
necessarily a local character. More precisely, single roots of this equation
such that locally a) $Im(\omega )<0$, b) $Im(\omega )=0$ or c)$\ Im(\omega
)>0$ will be referred to as (locally) stable, marginally stable and unstable
respectively. Correspondingly, the perturbation $\left( \delta g^{\mu \nu
}(s),\delta \pi _{\mu \nu }(s)\right) $ will be classified to be locally a)
decaying, b) oscillatory; c) growing. Therefore, the equilibrium solution $%
\widehat{g}(r)$ will be denoted as a) locally stable, b) locally marginally
stable and c) locally unstable respectively if:

a) all roots of $\omega $ are locally stable: namely\ for them $Im(\omega
)<0;$

b) there is at least one root of $\omega $ which is locally marginally
stable, namely such that $Im(\omega )=0$;

c) there is at least one locally unstable root of $\omega ,$ namely such
that $Im(\omega )>0.$

To investigate the stability problem we consider the vacuum configuration in
which $T_{\mu \nu }\equiv 0$, but still $\Lambda \neq 0$,\ so that $\widehat{%
R}_{\mu \nu }=\Lambda \widehat{g}_{\mu \nu }$ and $\widehat{R}=4\Lambda $ is
the constant Ricci scalar. Then Eq.(\ref{DISPERSION RELATION}) yields the
dispersion relation in the explicit form%
\begin{equation}
\left( -\left[ \frac{\omega }{c}-K\right] ^{2}-5\sigma \Lambda \right)
^{2}-4\left( -\sigma \Lambda \right) ^{2}=0,  \label{DISPERSION RELATION-2}
\end{equation}%
which delivers the roots%
\begin{equation}
\frac{\omega }{c}-K=\left\{
\begin{array}{c}
\pm \sqrt{-7\sigma \Lambda } \\
\pm \sqrt{-3\sigma \Lambda }%
\end{array}%
\right. .
\end{equation}%
Therefore, two possible alternatives can be distinguished in which
respectively:

A) \emph{First case:} $-\sigma \Lambda \geq 0$. Then the equilibrium
solution $\widehat{g}(r)$ is marginally stable since all roots of the
dispersion relation (\ref{DISPERSION RELATION-2}) have vanishing imaginary
part.

B) \emph{Second case: }$-\sigma \Lambda <0$. Then $\widehat{g}(r)$ is
necessarily unstable (there exists always an unstable root of the same
equation (\ref{DISPERSION RELATION-2})).

Hence, both for the case of de Sitter and anti-de Sitter space times ($%
\Lambda $\ respectively $>0$\ or $<0$) the possible occurrence of stability
or instability depends on the multiplicative gauge parameter $\sigma $\
appearing in the definition of the effective potential density $V$\ (see the
second equation given above in (\ref{V_R-N})). However, a
physically-admissible Hamiltonian theory of GR should predict stable
solutions, i.e., which are structurally stable in the sense indicated above.
This should occur in principle not just for special realizations of the
background space-time but - at least in the case of vacuum \textbf{- }\emph{%
for arbitrary vacuum background space-times }$\left( \mathbf{Q}^{4},\widehat{%
g}(r)\right) $. In particular, if $\Lambda >0$ - as most frequently invoked
in the literature (see for example \cite{Winberg2000,Carroll2004}) - this
happens provided the gauge factor $\sigma $\ is uniquely identified with $%
\sigma =-1$. Although the rigorous validity of such a choice still remains a
mere conjecture at this point, its\ full justification should ultimately
emerge from quantum theory. Nevertheless, the stability property pointed out
here can be viewed as prerequisite for the consistent development of a
covariant quantum gravity theory. For this reason the issue will be further
discussed in Part 2.

\section{6 - Conclusions}

A common fundamental theoretical aspect laying at the foundation of both
General Relativity (GR) and classical field theory is the variational
character of the fundamental dynamical laws which identify these
disciplines. This concerns both the representation of the Einstein field
equations and the covariant dynamics of classical fields as well as of
discrete (e.g., test particles) or continuum systems in curved space-time.
Issues related to the variational formulation of the Einstein equations have
been treated in Refs.\cite{noi1,noi2}, where the existence of a new type of
Lagrangian and Hamiltonian variational approaches has been identified in
terms of synchronous variational principles realized in the framework of the
DeDonder-Weyl formalism. As shown in Ref.\cite{noi2}, this leads to the
realization of a manifestly covariant Hamiltonian theory for the Einstein
equations.

In this paper new aspects of the Hamiltonian structure of GR have been
displayed which is referred to here as CCG-theory.

In particular, we have shown that a reduced-dimensional realization of the
continuum Hamiltonian theory for the Einstein field equations, denoted here
as CHS (Classical Hamiltonian Structure of GR), actually exists\ in which
both generalized coordinates and corresponding conjugate momenta are
realized by means of 2nd-order $4-$tensors. The virtue of such an approach
lies precisely in its general validity. This means in fact that the same
Hamiltonian structure holds for arbitrary particular solutions of the
Einstein field equations and arbitrary realizations of the external source
terms appearing in the variational potential density. As a result, a causal
form has been obtained for the corresponding continuum Hamilton equations\
by introducing a suitable Lagrangian parametrization prescribed in terms of
the proper-time\emph{\ }$s$ defined along field geodetics of the curved
space-time. As a result, the same equations are cast in the equivalent form
of an initial-value problem for suitable canonical evolution equations,
referred to here as GR-Hamilton equations.\emph{\ }This provides a physical
interpretation for the reduced Hamiltonian theory, according to which an
arbitrary initial canonical state is dynamically advanced by means of the
canonical flow generated by the same Hamilton equations.

Given validity to such Hamiltonian theory, the case of canonical
transformations which generate the flow corresponding to the continuum
canonical equations has been considered. This has been obtained by
introducing the appropriate mixed-variable generating function - the
so-called Hamilton principal function - and by developing the corresponding
Hamilton-Jacobi theory in manifestly-covariant form. As a result, the same
generating function has been shown to obey a $4-$scalar continuum
Hamilton-Jacobi equation which has been proved to be equivalent to the
corresponding canonical evolution equations of the Hamiltonian theory. The
global prescription and regularity of the Hamiltonian structure have been
analyzed and the gauge transformation properties of the reduced Hamiltonian
density $H_{R}$\ have been pointed out.

Finally, as an application of the Hamiltonian formulation developed here,
the structural stability of the Hamiltonian theory has been investigated, a
feature which is required for a consistent development of a corresponding
quantum theory of GR based on the same canonical representation. In this
paper we have studied the stability of\ perturbed fields which realize
particular solutions of the GR-Hamilton equations with respect to stationary
solutions, \textit{i.e.}, metric tensor solutions of the Einstein field
equations. The case of background vacuum solutions having vanishing
stress-energy tensor and non-null cosmological constant has been analyzed,
determining the conditions for the occurrence of stable and unstable roots,
adopting an eikonal representation for the perturbed fields.

These conclusions highlight the major key features of the reduced
Hamiltonian theory and corresponding Hamilton-Jacobi equation determined
here. The new theory, besides being a mandatory prerequisite for the
covariant theory of quantum gravity to be established in Part 2, is believed
to be susceptible of applications to physics and astrophysics-related
problems and to provide at the same time new insight in\ the axiomatic
foundations of General Relativity.

\section{Acknowledgments}

Work developed within the research projects of the Czech Science Foundation
GA\v{C}R grant No. 14-07753P (C.C.) and the Albert Einstein Center for
Gravitation and Astrophysics, Czech Science Foundation No. 14-37086G (M.T.).

\end{document}